\documentclass[showpacs,twocolumn,amsmath,amssymb,pra,superscriptaddress,nofootinbib]{revtex4}

\usepackage{graphicx}

\begin{document}
\title{Strong monotonicity in mixed-state entanglement manipulation}
\author{Satoshi Ishizaka}
\affiliation{
Fundamental Research Laboratories, NEC Corporation, 
34 Miyukigaoka, Tsukuba, 305-8501, Japan}
\affiliation{
PRESTO, Japan Science and Technology Agency,
4-1-8 Honcho Kawaguchi, 332-0012, Japan}
\date{\today}
\begin{abstract}
A strong entanglement monotone, which never increases under local operations
and classical communications (LOCC), restricts quantum entanglement
manipulation more strongly than the usual monotone since the usual one does
not increase on average under LOCC.
We propose new strong monotones in mixed-state entanglement manipulation under
LOCC.
These are related to the decomposability and 1-positivity of an operator
constructed from a quantum state, and reveal geometrical characteristics of
entangled states.
These are lower bounded by the negativity or generalized robustness of
entanglement.
\end{abstract}
\maketitle
\section{Introduction}
\label{sec: Introduction}
%
It is a key concept for quantum information science that distant parties can
manipulate quantum entanglement by local operations and classical communication
(LOCC).
However, the entanglement manipulation suffers some fundamental
restrictions: LOCC cannot create entanglement \cite{Yang05a} and
LOCC cannot increase the total amount of entanglement.
This monotonicity is characterized by mathematical functions
$E(\sigma)$, called entanglement monotones \cite{Vidal00a}.
When a quantum state $\varrho_i$ is obtained from $\sigma$ with probability
$p_i\!>\!0$ by LOCC ($i$ indexes the multiple outcomes),
the functions satisfy
\begin{equation}
E(\sigma)\!\ge\!\sum_i p_i E(\varrho_i),
\end{equation}
and therefore $E(\sigma)$ does not increase {\it on average} under LOCC.
Many such monotones have been proposed such as entanglement measures
(e.g. entanglement cost \cite{Bennett96a},
distillable entanglement \cite{Bennett96a},
and relative entropy of entanglement \cite{Vedral97a}),
negativity \cite{Vidal02a,Plenio05a},
robustness of entanglement \cite{Vidal99b,Steiner03a},
best separable approximation (BSA) measure \cite{Karnas01a,Lewenstein98a},
and so on \cite{PlenioVirmani05a}.
\par
%
On the other hand, there exists a much stronger restriction in the entanglement
manipulation: the Schmidt number \cite{Terhal00b},
which is a general extension of the Schmidt rank to mixed states, 
cannot increase even with an infinitesimally small probability.
This type of restriction, called strong monotonicity here, may be characterized
by strong monotone functions $M(\sigma)$ which satisfy
\begin{equation}
M(\sigma)\!\ge\! M(\varrho_i) \hbox{~~for all $\varrho_i$}.
\end{equation}
Namely $M(\sigma)$ {\it never} increases under LOCC.
Note that the strong monotones are generally discontinuous functions
as explicitly shown later.
Concerning the conversion between bipartite pure states, the Schmidt number
is a unique strong monotone
(the conversion is impossible when the Schmidt rank of the target state is
larger than that of the initial state, but otherwise the conversion
is possible with nonzero probability \cite{Vidal99a,Dur00b}).
Positive partial transpose preserving (PPT-preserving) operations
can overcome the monotonicity of the Schmidt number,
and therefore all pure entangled states become convertible under
PPT-preserving operations \cite{Ishizaka04b,Ishizaka05a}.
Here, a map $\Lambda$ is called PPT-preserving operations \cite{Rains99b} 
when both $\Lambda$ and $\Gamma\!\circ\!\Lambda\!\circ\!\Gamma$ is a
completely positive (CP) map with $\Gamma$ being a map of the partial
transpose \cite{Peres96a}.
On the other hand, the manipulation of mixed-state
entanglement still suffers some restriction even under PPT-preserving
operations in single-copy settings \cite{Ishizaka04b,Ishizaka05a}.
This implies that there certainly exists strong monotonicity independent
on the Schmidt number in mixed-state entanglement manipulation.
\par
%
In this paper, we propose new strong monotones which are related to the
decomposability and 1-positivity of an operator constructed from a quantum
state.
Here, an operator $Z$ is called decomposable when $Z$ is written as
$Z\!=\!X\!+\!Y^\Gamma$ with $X,Y\!\ge\!0$,
and $Z$ is called 1-positive when $\langle ef|Z|ef \rangle\!\ge\!0$ for
every product states $|ef\rangle$
(see e.g. 
\cite{Horodecki96a,Lewenstein00a,Sanpera01a,Kraus02a,Ha03a,Clarisse05a}
for the relation between entanglement and decomposability or 1-positivity).
The singlet fraction and negativity maximized over (stochastic) LOCC are also
strong monotones, but strong monotones studied in this paper are slightly
different from those and reveal geometrical characteristics of entangled
states.
These are lower bounded by the negativity or generalized
robustness of entanglement.
\par
\section{Strong monotone $M_1$}
\label{sec: Strong monotone M_1}
%
The first strong monotone we propose is the following:
\par
{Theorem 1:}
{\it
Let $\sigma^\Gamma\!=\!P-Q$ be the Jordan decomposition (orthogonal
decomposition) of $\sigma^\Gamma$
and hence $P\!\ge\!0$, $Q\!\ge\!0$, and $PQ\!=\!0$.
The function $M_1(\sigma)$, which is defined as the minimal x
subject that $\sigma\!-\!(1\!-\!x)P^\Gamma$ is decomposable, is a strong
monotone, i.e. $M_1(\sigma)$ never increases under LOCC (and even under
PPT-preserving operations).
}
\par
%
Note that $\sigma\!-\!(1\!-\!x)P^\Gamma$ is always decomposable for $x\!=\!1$,
and hence $M_1(\sigma)\!\le\!1$.
Before proving the above theorem, let us show explicit examples of
$M_1(\sigma)$ for several important classes of states.
\par
(i) {\it Separable states:} For every PPT states ($\sigma^\Gamma\!\ge\!0$),
we have $P^\Gamma\!=\!\sigma$ and hence
$\sigma\!-\!(1\!-\!x)P^\Gamma\!=\!x\sigma$ which is decomposable only for
$x\!\ge\!0$.
Since all separable states are PPT states, $M_1(\sigma)\!=\!0$ for every
separable $\sigma$.
\par
(ii) {\it Entangled pure states:}
Let 
\begin{equation}
|\phi_d^+\rangle=\frac{1}{\sqrt{d}}\sum_{i=0}^{d-1}|ii\rangle
\end{equation}
be a maximally entangled
state on ${\mathbb C}^d\!\otimes\!{\mathbb C}^d$ and
$P^+_d\!\equiv\!|\phi_d^+\rangle\langle\phi_d^+|$.
When $\sigma\!=\!P^+_d$, $P\!=\!P_d^S/d$ where $P_d^S$ is the projector
onto the symmetric subspace on ${\mathbb C}^d\!\otimes\!{\mathbb C}^d$.
Any decomposable operator $Z$ must satisfy
$\langle ef|Z|ef\rangle\!\ge\!0$ for every product states $|ef\rangle$,
i.e. 1-positive, but 
\begin{equation}
\langle01|P^+_d-(1-x)\big(\frac{P_d^S}{d}\big)^\Gamma|01\rangle
=-\frac{1-x}{2d}.
\end{equation}
Therefore $P^+_d\!-\!(1\!-\!x)(P_d^S/d)^\Gamma$ cannot be decomposable for
$x\!<\!1$
and $M_1(P_d^+)\!=\!1$ independent on $d$.
It has been shown that an entangled $|\psi\rangle$ can be converted to
$P^+_r$ by LOCC where $r\!\ge\!2$ is the
Schmidt number of $|\psi\rangle$ \cite{Vidal99a}.
Since $M_1$ never increases under LOCC and $M_1\!\le\!1$,
we have $M_1(|\psi\rangle)\!=\!1$ for every entangled $|\psi\rangle$.
Therefore, $M_1$ does not distinguish entangled pure states at all.
This property however is desirable for the purpose of this paper
that is to study strong monotonicity independent on the Schmidt number.
As mentioned before, the Schmidt number is a unique strong monotone
concerning the conversion between pure states.
Therefore, the strong monotones independent on the Schmidt number should not
distinguish entangled pure states.
\par
%
It should be noted that $M_1(|ef\rangle)\!=\!0$ as shown in (i) but 
in the close vicinity of $|ef\rangle$ there always exists a partially
entangled pure state $|\psi\rangle$ for which $M_1(|\psi\rangle)\!=\!1$.
As a result, it is found that $M_1(\sigma)$ is a discontinuous function.
\par
%
According to the above examples (i) and (ii), the following is concluded:
\par
{Corollary 1:}
{\it
If $\sigma$ is single-copy distillable under LOCC (or under PPT-preserving operations), then $M_1(\sigma)\!=\!1$.
}
\par
%
(iii) {\it Antisymmetric Werner states:}
For an antisymmetric Werner state of
$\sigma^A_d\!=\!2/(d^2\!-\!d)P_d^A$,
$P\!=\!(\openone_d\!-\!P_d^+)/(d^2\!-\!d)$ where
$\openone_d$ and $P_d^A$ is the identity on ${\mathbb C}^d\!\otimes\!{\mathbb C}^d$
and projector onto the antisymmetric subspace on
${\mathbb C}^d\!\otimes\!{\mathbb C}^d$, respectively.
Then,
\begin{equation}
\langle00|\sigma^A_d-(1-x)\big(\frac{\openone_d-P_d^+}{d^2-d}\big)^\Gamma|00\rangle
=-\frac{1-x}{d^2},
\end{equation}
which cannot be decomposable for $x\!<\!1$, and $M_1(\sigma^A_d)\!=\!1$.
\par
(iv) {\it Convex combination of $\sigma_0$ and $P_0^\Gamma$:}
Let $\sigma_0^\Gamma\!=\!P_0-Q_0$ be
the Jordan decomposition.
For the state of
\begin{equation}
\sigma=\sigma_0+\lambda P_0^\Gamma,
\label{eq: convex combination}
\end{equation}
$\sigma^\Gamma\!=\!(1+\lambda)P_0\!-\!Q_0$
(here $\sigma$ is not normalized but $M_1(\sigma)$ does not depend on the
normalization).
As a result,
\begin{eqnarray}
\sigma-(1-x)P^\Gamma&\!\!\!=\!\!\!&\sigma_0+\lambda P_0^\Gamma
-(1-x)(1+\lambda)P_0^\Gamma \cr
&\!\!\!=\!\!\!&\sigma_0-\big[1-x(1+\lambda)]P_0^\Gamma,
\end{eqnarray}
and therefore
\begin{equation}
M_1(\sigma)=\frac{M_1(\sigma_0)}{1+\lambda}
\end{equation}
for the state of Eq.\ (\ref{eq: convex combination}).
An entangled isotropic state 
\begin{equation}
\sigma_I=\eta P_d^++(1-\eta)\frac{\openone_d-P_d^+}{d^2-1},
\label{eq: Isotropic state}
\end{equation}
where $\eta\!>\!1/d$, can be rewritten as
$\sigma_I\!\propto\!\sigma_0\!+\!\frac{2d(1-\eta)}{(d\eta-1)(d+1)}P_0^\Gamma$
with $\sigma_0\!=\!P_d^+$ [correspondingly $P_0^\Gamma\!=\!(\openone_d\!+\!dP_d^+)/(2d)$], and therefore
\begin{equation}
M_1(\sigma_I)=\frac{(d\eta-1)(d+1)}{(d\eta+1)(d-1)}.
\label{eq: monotone for an isotropic state}
\end{equation}
Similarly, for an entangled Werner state
\begin{equation}
\sigma_W=\mu \frac{2}{d^2-d}P_d^A + (1-\mu)\frac{2}{d^2+2}P_d^S,
\end{equation}
where $\mu\!>\!1/2$, putting $\sigma_0\!=\!\sigma^A_d$ we have 
\begin{equation}
M_1(\sigma_W)=\frac{(2\mu-1)(d+1)}{2\mu+d-1}.
\label{eq: monotone for a Werner state}
\end{equation}
\par
%
It has been mentioned that $\eta$ of an isotropic state and $\mu$ of a Werner
state cannot increase under LOCC \cite{Masanes05a}.
This can be confirmed by $M_1(\sigma_I)$ and $M_1(\sigma_W)$,
and moreover it is found that this is the case even under PPT-preserving
operations,
since these functions are monotonic with respect to $\eta$ and $\mu$,
respectively, and these never increase under LOCC and even under PPT-preserving
operations.
\par
%
Equating Eq.\ (\ref{eq: monotone for an isotropic state}) and
Eq.\ (\ref{eq: monotone for a Werner state}), we have a relation:
\begin{equation}
\mu=\frac{(d-1)\eta}{(d-2)\eta+1},
\label{eq: mu and eta}
\end{equation}
and therefore the reversible conversion of
$\sigma_W\!\leftrightarrow\!\sigma_I$ is not
prohibited by $M_1$ if $\mu$ and $\eta$ satisfy Eq.\ (\ref{eq: mu and eta}).
Indeed, this reversible conversion is possible by PPT-preserving operations,
whose trace non-preserving maps are
\begin{eqnarray*}
&&\textstyle
\sigma_W\!\rightarrow\!\sigma_I:
\Lambda(X)=(\hbox{tr}XP_d^A)P_d^+
+(\hbox{tr}XP_d^S)\frac{\openone_d-P_d^+}{d+1}, \cr
&&\textstyle
\sigma_I\!\rightarrow\!\sigma_W:
\Lambda(X)=(\hbox{tr}XP_d^+)P_d^A
+(\hbox{tr}X(\openone_d\!-\!P_d^+))\frac{P_d^S}{d+1}.
\end{eqnarray*}
It should be noted however that the conversion of
$\sigma_W\!\rightarrow\!\sigma_I$ under LOCC suffers the strong monotonicity
due to the Schmidt number because it is 2 for an entangled $\sigma_W$ but
larger than 2 for $\sigma_I$ when $\eta\!>\!2/d$ ($d\!>\!2$) \cite{Terhal00b}.
\par
%
(v) {\it Two-qubit states:}
Let us consider an entangled Bell diagonal state:
\begin{equation}
\sigma_B=\sum_{i=0}^{3} p_i |e_i\rangle\langle e_i|, \,\,\,
(p_0 \ge p_1 \ge p_2 \ge p_3),
\end{equation}
where $p_0\!>\!1/2$ so that $\sigma_B$ is entangled.
The Bell basis is chosen as 
$|e_i\rangle\!=\!\{i|\psi^-\rangle,|\psi^+\rangle,|\phi^-\rangle,i|\phi^+\rangle\}$
with $|\phi^{\pm}\rangle\!=\!(|00\rangle\!\pm\!|11\rangle)/\sqrt{2}$ and
$|\psi^{\pm}\rangle\!=\!(|01\rangle\!\pm\!|10\rangle)/\sqrt{2}$, which
is the magic basis and hence 
$|\tilde e_i\rangle\!=\!(\sigma_2\otimes\sigma_2)|e_i^*\rangle\!=\!
|e_i\rangle$ \cite{Bennett96a,Wootters98a}.
Then we have
\begin{eqnarray}
\lefteqn{\sigma_B-(1-x)P^\Gamma=
\big[p_0-\frac{(1-x)(2p_0-1)}{4}\big]|e_0\rangle\langle e_0|}\quad\quad\quad\cr
&&+\sum_{i=1}^{3}\!\big[p_i\!-\!\frac{(1\!-\!x)(2p_0\!-\!1\!+\!4p_i)}{4}\big]|e_i\rangle\langle e_i|. \quad
\label{eq: Bell diagonal}
\end{eqnarray}
For the decomposability of such a Bell diagonal operator, the following is
useful:
\par
%
{Lemma 1:}
{\it
An operator $A\!=\!\sum_{i=0}^{3} a_i |e_i\rangle\langle e_i|$ on
${\mathbb C}^2\!\otimes\!{\mathbb C}^2$,
where $a_i\!\ge\!a_{i+1}$, is decomposable (and 1-positive) if and only if
$a_2\!+\!a_3\!\ge\!0$.
}
\par
{\it Proof:} When $A$ on ${\mathbb C}^2\!\otimes\!{\mathbb C}^2$
is expressed
as $A\!=\!(I\!\otimes\!\Theta)P_2^+$ with $\Theta$ being a map, the following
four statements are equivalent
\cite{Stormer63a,Woronowicz76a}: (a) $A$ is decomposable, (b) $\Theta$ is a
decomposable positive map, (c) $\Theta$ is a
positive map, and (d) $A$ is 1-positive.
Since $|e_i\rangle$ is an orthogonal set, any pure state is expanded as
$|\psi\rangle=\sum_{i=0}^{3}\lambda_i|e_i\rangle$ with $\sum_i|\lambda_i|^2\!=\!1$.
So that $|\psi\rangle$ is a product state,
$\langle\tilde \psi|\psi\rangle\!=\!\sum \lambda_i^2\!=\!0$ \cite{Bennett96a}.
When the real and imaginary part of $\lambda_i$ is $r_i$ and $c_i$,
respectively, the above
two conditions are written as $\sum_i r_i^2\!=\!\sum_i c_i^2\!=\!1/2$
and $\sum_i r_i c_i=0$.
Therefore, the four dimensional real vectors $\vec r$ and $\vec c$, whose
elements are $r_i$ and $c_i$, respectively, satisfy
$|\vec r|^2\!=\!|\vec c|^2\!=\!1/2$ and $\vec r\cdot \vec c\!=\!0$, and
it is easy to see $r_3^2\!+\!c_3^2\!\le\!1/2$.
Then,
\begin{eqnarray*}
\langle \psi |A|\psi\rangle&\!\!\!=\!\!\!&\sum_{i=0}^3|\lambda_i|^2a_i
\ge \sum_{i=0}^2|\lambda_i|^2a_2+|\lambda_3|^2a_3 \cr
&\!\!\!=\!\!\!&a_2-(a_2-a_3)(r_3^2+c_3^2)\ge \frac{1}{2}(a_2+a_3),
\end{eqnarray*}
and therefore if $a_2\!+\!a_3\!\ge\!0$ then $\langle ef |A|ef\rangle\!\ge\!0$
for every $|ef\rangle$.
Conversely, if $a_2\!+\!a_3\!<\!0$ then
$\langle 00 |A|00\rangle\!<\!0$.
\hfill \mbox{\vrule width .6em height .6em}
\par
Using this lemma, it is found that Eq.\ (\ref{eq: Bell diagonal}) is
decomposable if and only if $x\!\ge\!(2p_0\!-\!1)/(1\!-\!2p_1)$, and 
\begin{equation}
M_1(\sigma_B)=\frac{2p_0-1}{1-2p_1}
\label{eq: monotone for Bell}
\end{equation}
for an entangled $\sigma_B$.
It has been shown that $p_0$ of the entangled Bell diagonal state cannot
increase under LOCC \cite{Verstraete03b}.
By Eq.\ (\ref{eq: monotone for Bell}), it is found that $p_1$ also
cannot increase unless $p_0$ is decreased.
It has been shown further that almost all entangled two-qubit states can be
converted to
Bell diagonal states by LOCC in a reversible fashion \cite{Verstraete01a}.
Such reversible LOCC does not change $M_1$, and hence $M_1(\sigma)$ for
such the entangled two-qubit state $\sigma$ agrees with
Eq.\ (\ref{eq: monotone for Bell}) of the converted $\sigma_B$.
\par
Note that Eq.\ (\ref{eq: monotone for Bell}) is
equal to 1 for $p_1\!=\!1\!-\!p_0$, and therefore $M_1(\sigma)\!=\!1$ for
every entangled two-qubit states of rank-2.
However, two-qubit states of rank-2 are not single-copy distillable
\cite{Kent98a} (though a special state is quasi distillable 
\cite{Horodecki99a,Verstraete01a}) under LOCC
and even under PPT-preserving operations \cite{Ishizaka04b,Ishizaka05a}.
Therefore the converse of the corollary 1 does not hold in general.
\par
%
\section{A family of strong monotones}
\label{sec: A family of strong monotones}
%
Let us now prove the theorem 1.
Our starting point is the following function:
\begin{equation}
M(\sigma)=\min_{\sigma_\pm\in{\cal C}_\pm} \sup_{\Omega \in {\cal L}_\sigma}
\frac{\hbox{tr}\Omega (\sigma_-)}
{\hbox{tr}\Omega (\sigma_+)},
\label{eq: function M}
\end{equation}
where the minimization is performed over all possible decompositions
of $\sigma\!=\!\sigma_+\!-\!\sigma_-$ such as $\sigma_+\!\in\!{\cal C}_+$ and
$\sigma_-\!\in\!{\cal C}_-$.
Note that $\sigma_\pm$ (unnormalized) are not necessarily positive,
and the sets ${\cal C}_\pm$ are specified later.
The supremum is taken for all possible operations $\Omega$
that belong to some operational class ${\cal L}_\sigma$, which may depend
on $\sigma$.
Then, suppose that $\sigma$ is converted to $\varrho$ by
LOCC or PPT-preserving operations with nonzero
probability $p$, and hence there exists a PPT-preserving map
$\Lambda_{\sigma\rightarrow\varrho}$ such that
$\Lambda_{\sigma\rightarrow\varrho}(\sigma)\!=\!p\varrho$
(LOCC is also PPT-preserving).
Moreover suppose that the sets ${\cal C}_\pm$ have been chosen such that
$(1/p)\Lambda_{\sigma\rightarrow\varrho}(\sigma_\pm)\!\equiv\!\varrho_\pm
\!\in\!{\cal C}_\pm$, and
suppose that ${\cal L}_\sigma$ has been chosen such that 
$\Omega\!\circ\!\Lambda_{\sigma\rightarrow\varrho}\!\in\!{\cal L}_\sigma$
for every $\Omega\!\in\!{\cal L}_\varrho$.
Under these assumptions, the function $M(\sigma)$ is indeed a strong monotone
because
\begin{eqnarray}
M(\sigma)&\!\!\!=\!\!\!&\min_{\sigma_\pm\in{\cal C}_\pm}
\sup_{\Omega \in {\cal L}_\sigma}
\frac{\hbox{tr}\Omega (\sigma_-)}{\hbox{tr}\Omega (\sigma_+)} \cr
&\!\!\!\ge\!\!\!&\min_{\sigma_\pm\in{\cal C}_\pm}
\sup_{\Omega \in {\cal L}_\varrho}
\frac{\hbox{tr}\Omega \circ\Lambda_{\sigma\rightarrow\varrho} (\sigma_-)}
{\hbox{tr}\Omega \circ \Lambda_{\sigma\rightarrow\varrho} (\sigma_+)} \cr
&\!\!\!=\!\!\!&\min_{\sigma_\pm\in{\cal C}_\pm}
\sup_{\Omega \in {\cal L}_\varrho}
\frac{\hbox{tr}\Omega (\varrho_-)}{\hbox{tr}\Omega (\varrho_+)} \cr
&\!\!\!\ge\!\!\!&\min_{\varrho_\pm\in{\cal C}_\pm}
\sup_{\Omega \in {\cal L}_\varrho}
\frac{\hbox{tr}\Omega (\varrho_-)}{\hbox{tr}\Omega (\varrho_+)}=M(\varrho),
\label{eq: strong monotonicity}
\end{eqnarray}
and $M$ never increases under the conversion of $\sigma\!\rightarrow\!\varrho$
if it is possible with nonzero probability.
Note that the function $M$ is neither convex nor concave in general.
\par
%
Then, let us consider the case where ${\cal C}_\pm\!=\!\{A|A^\Gamma\!\ge\!0\}$.
Namely, the minimization in Eq.\ (\ref{eq: function M}) is performed
over all possible decomposition such as 
$\sigma^\Gamma\!=\!a_+\!-\!a_-$ with $a_+\!\ge\!0$ and $a_-\!\ge\!0$.
This is the same decomposition introduced in \cite{Vidal02a} 
as the minimization problem for the negativity
(see also \cite{PlenioVirmani05a}).
Since $\Lambda_{\sigma\rightarrow\varrho}$ is PPT-preserving,
$\Gamma\!\circ\!\Lambda_{\sigma\rightarrow\varrho}\!\circ\!\Gamma$ is a CP map.
Therefore
$\varrho_\pm^\Gamma\!=\!(1/p)\Gamma\!\circ\!
\Lambda_{\sigma\rightarrow\varrho}\!\circ\!\Gamma(a_\pm)\!\ge\!0$
and $\varrho_\pm\!\in\!{\cal C}_\pm$ is satisfied
\cite{Vidal02a}.
The explicit form of the function $M(\sigma)$ in this case, denoted by
$M_1(\sigma)$, is
\begin{equation}
M_1(\sigma)=
\min_{a_\pm\ge0} \sup_{\Omega \in {\cal L}_\sigma}
\frac{\hbox{tr}\Omega (a_-^\Gamma)}
{\hbox{tr}\Omega (a_+^\Gamma)}.
\label{eq: monotone 1}
\end{equation}
where ${\cal L}_\sigma$ is chosen as a set of PPT-preserving operations
restricted to $\hbox{tr}\Omega(\sigma)\!>\!0$ (note that the optimization
is supremum considering $\hbox{tr}\Omega(\sigma)\!\rightarrow\!0$).
For this choice of ${\cal L}_\sigma$,
$\Omega\!\circ\!\Lambda_{\sigma\rightarrow\varrho}\!\in\!
{\cal L_\sigma}$ for every $\Omega\!\in\!{\cal L_\varrho}$ and hence
the strong monotonicity Eq.\ (\ref{eq: strong monotonicity}) holds.
Here, suppose that there exists a nonzero positive operator $a\!\ge\!0$ such
that $a'_\pm\!\equiv\!a_\pm\!-\!a\!\ge\!0$
for some decomposition of $\sigma^\Gamma\!=\!a_+\!-\!a_-$.
Then,
\begin{equation}
\max_\Omega 
\frac{\hbox{tr}\Omega (a^\Gamma_-)}
{\hbox{tr}\Omega (a_+^\Gamma)}
=
\max_\Omega 
\frac{\hbox{tr}\Omega (a'^\Gamma_-)
\!+\!\hbox{tr}\Omega (a^\Gamma)}
{\hbox{tr}\Omega(a'^\Gamma_+)
\!+\!\hbox{tr}\Omega(a^\Gamma)}
\ge
\max_\Omega 
\frac{\hbox{tr}\Omega (a'^\Gamma_-)}
{\hbox{tr}\Omega(a'^\Gamma_+)},
\nonumber
\end{equation}
where $\hbox{tr}\Omega(a^\Gamma)\!=\!
\hbox{tr}[\Gamma\!\circ\!\Omega\!\circ\!\Gamma(a)]\!\ge\!0$ and
$\hbox{tr}\Omega(a'^\Gamma_+)=\hbox{tr}\Omega(\sigma)\!+\!\hbox{tr}\Omega(a'^\Gamma_-)\!\ge\!\hbox{tr}\Omega(a'^\Gamma_-)$ were used.
As a result, it is found that the minimization in Eq.\ (\ref{eq: monotone 1})
is reached when $a_\pm$ are orthogonal to each other and hence
\begin{equation}
M_1(\sigma)=\max_{\Omega\in{\cal L}_\sigma}\frac{\hbox{tr}\Omega(Q^\Gamma)}{\hbox{tr}\Omega(P^\Gamma)}
\label{eq: monotone 1b}
\end{equation}
with $\sigma^\Gamma\!=\!P\!-\!Q$ being the Jordan decomposition.
Moreover, it can be assumed that $\Omega\!=\!{\cal R}\!\circ\!\Omega$
where 
\begin{equation}
{\cal R}(X)=\int dUdV (U \otimes V)X(U \otimes V)^\dagger
\end{equation}
is the random application of local unitary transformations.
All such PPT-preserving operations have the form of
\begin{equation}
\Omega(X)=(\hbox{tr}XA)\openone
\end{equation}
with $A\!\ge\!0$ and $A^\Gamma\!\ge\!0$ \cite{Cirac01a}.
Moreover, since $\Omega$ is restricted to $\hbox{tr}\Omega(\sigma)\!>\!0$
and hence
$\hbox{tr}\Omega(P^\Gamma)=\hbox{tr}\Omega(\sigma)\!+\!\hbox{tr}\Omega(Q^\Gamma)\!>\!0$,
we can assume $\hbox{tr}P^\Gamma A\!=\!1$ without loss of generality.
Then, $M_1(\sigma)$ is reduced to
$M_1(\sigma)\!=\!\max_{A} \hbox{tr}(Q^\Gamma A)$ subject that
\begin{equation}
A\ge 0, \,\,\, A^\Gamma\ge0, \,\,\, \hbox{tr}P^\Gamma A=1.
\label{eq: primary constraints}
\end{equation}
This is a convex optimization problem, whose optimal value coincides with
the optimal value of the dual problem \cite{Boyd04a}.
Since
\begin{eqnarray}
\hbox{tr}Q^\Gamma A&\!\!\!=\!\!\!&
-\hbox{tr}Y A^\Gamma -x (1-\hbox{tr}P^\Gamma A) \cr
&&-\hbox{tr} A(-Q^\Gamma+xP^\Gamma-Y^\Gamma)+x,
\end{eqnarray}
where $x$ is a Lagrange multiplier, the dual problem is
$M_1(\sigma)\!=\!\min x$
subject to the constraints of $Y\!\ge\!0$ and
\begin{equation}
X\equiv-Q^\Gamma+xP^\Gamma-Y^\Gamma=\sigma-(1-x)P^\Gamma-Y^\Gamma\ge0.
\end{equation}
These constraints can be read as
$\sigma\!-\!(1\!-\!x)P^\Gamma\!=\!X\!+\!Y^\Gamma$ with $X,Y\!\ge\!0$,
and consequently the theorem 1 is obtained.
\par
\begin{figure}
\centerline{\scalebox{0.45}[0.45]{\includegraphics{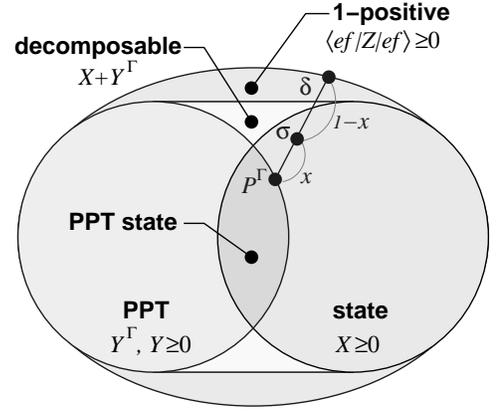}}}
\caption{
The set of positive operators (unnormalized states) and
set of PPT operators are schematically shown as two circles.
The intersection of the two circles corresponds to the set of (unnormalized)
PPT states.
The set of decomposable operators corresponds to the convex cone of the two
circles.
Moreover, the set of 1-positive operators embodies the set of decomposable
operators.
When $\sigma^\Gamma\!=\!P\!-\!Q$ is the Jordan decomposition,
$P^\Gamma$ is located on the edge of the set of PPT operators.
$P^\Gamma$ is an unnormalized PPT state for many classes of states, but
sometimes $P^\Gamma$ is not a state (see \protect\cite{Ishizaka04a}).
On the other hand, $\delta\!\equiv\!\sigma\!-\!(1\!-\!x)P^\Gamma$ is located
on the edge of the set of 1-positive operators.
The ratio of the interior division $x$ corresponds to the strong monotone
$M_1^{sep}(\sigma)$.
When $\delta$ is located on the edge of the set of decomposable operators,
the ratio of the interior division $x$ corresponds to the strong monotone
$M_1(\sigma)$.
}
\label{fig: Geometrical structure}
\end{figure}

It is obvious that
$M_1(\sigma)$ is a strong monotone under PPT-preserving operations
because ${\cal L}_\sigma$ was chosen as a set of PPT-preserving operations.
Then, let us consider another function $M_1^{sep}(\sigma)$ for which
${\cal L}_\sigma$ in Eq.\ (\ref{eq: monotone 1}) is replaced by the set of
LOCC restricted to $\hbox{tr}\Omega(\sigma)\!>\!0$.
By this, 
since the set of stochastic LOCC coincides with the set of stochastic
separable operations, the constraints of Eq.\ (\ref{eq: primary constraints})
become \cite{Cirac01a}
\begin{equation}
A\ge 0, \,\,\, A\hbox{~is separable}, \,\,\, \hbox{tr}P^\Gamma A=1.
\end{equation}
Following an idea of \cite{Brandao05a} and putting
\begin{equation}
A=\sum_i q_i |ef^{(i)}\rangle\langle ef^{(i)}|,
\end{equation}
with $q_i\!\ge\!0$, we have
\begin{eqnarray}
\hbox{tr}Q^\Gamma A&\!\!\!=\!\!\!&
-x \Big(1-\sum_i q_i\langle ef^{(i)}| P^\Gamma |ef^{(i)}\rangle\Big) \cr
&&-\sum_i q_i \langle ef^{(i)}|-Q^\Gamma+xP^\Gamma|ef^{(i)}\rangle +x.
\end{eqnarray}
Therefore, it is found that the corresponding
dual problem becomes $M_1^{sep}(\sigma)\!=\!\min x$ subject to
\begin{equation}
\langle ef|\sigma-(1-x)P^\Gamma|ef\rangle\ge 0 \hbox{~~for every $|ef\rangle$},
\label{eq: separable constraint}
\end{equation}
i.e. subject that $\sigma\!-\!(1\!-\!x)P^\Gamma$ is 1-positive.
This dual problem for $M_1^{sep}(\sigma)$ has a simple geometrical meaning
as well as $M_1(\sigma)$ as shown in Fig.\ \ref{fig: Geometrical structure}.
\par
%
Although $M_1^{sep}$ is a strong monotone only under LOCC (not under
PPT-preserving operations),
all examples (i)-(v) for $M_1$ also hold for $M_1^{sep}$ without any
modification, namely $M_1(\sigma)\!=\!M_1^{sep}(\sigma)$ for those classes
of states.
Note that when $Z\!\equiv\!\sigma\!-\!(1\!-\!x)P^\Gamma$ is expressed
as $Z\!=\!(F\!\otimes\!\Theta)P_d^+$,
if $\Theta$ is a non-decomposable positive map for some $x$, then
$M_1(\sigma)\!>\!M_1^{sep}(\sigma)$
(see also Fig.\ \ref{fig: Geometrical structure}).
Here $F$ is a map of the local filter that converts $\openone/d$
to $Z_A\!=\!\hbox{tr}_B Z$ ($Z_A$ must be positive so that $Z$ is 1-positive).
Note further that from Eq.\ (\ref{eq: monotone 1}),
$M_1(\sigma)$ and $M_1^{sep}(\sigma)$ is lower
bounded by the negativity $N(\sigma)\!=\!(\hbox{tr}|\sigma^\Gamma|\!-\!1)/2$ 
\cite{Vidal02a} as
\begin{equation}
M_1(\sigma)\ge M_1^{sep}(\sigma)\ge
\frac{\hbox{tr}Q}{\hbox{tr}P}=\frac{N(\sigma)}{1+N(\sigma)}.
\label{eq: Lower bound for M_1}
\end{equation}
\par
%
\section{Strong monotone $M_2$}
\label{sec: Strong monotone M_2}
Let us return to Eq.\ (\ref{eq: function M}) and consider
the case where ${\cal C}_+\!=\!\{A|A\!\ge\!0,A\hbox{~is separable}\}$ and
${\cal C}_-\!=\!\{A|A\!\ge\!0\}$,
namely the same decomposition for the minimization problem of the generalized
robustness of entanglement \cite{Vidal99b,Steiner03a}
(see also \cite{PlenioVirmani05a}).
Moreover, ${\cal L}_\sigma$ is chosen as a set of LOCC restricted to
$\hbox{tr}\Omega(\sigma)\!>\!0$.
When $\Lambda_{\sigma\rightarrow\varrho}$ is LOCC,
$\varrho_+\!=\!(1/p)\Lambda_{\sigma\rightarrow\varrho}(\sigma_+)$
is separable and $\varrho_\pm\!\in\!{\cal C}_\pm$ holds.
The explicit form of the function $M(\sigma)$ in this case, denoted by
$M_2^{sep}(\sigma)$, is
\begin{equation}
M_2^{sep}(\sigma)=\min_{\sigma_\pm} \sup_{\Omega \in {\cal L}_\sigma}
\frac{\hbox{tr}\Omega (\sigma_-)}
{\hbox{tr}\Omega (\sigma_+)}
\label{eq: monotone 2}
\end{equation}
with the constraints of $\sigma\!=\!\sigma_+\!-\!\sigma_-$,
$\sigma_\pm\!\ge\!0$, and $\sigma_+$ is separable.
Since $\hbox{tr}\Omega(\sigma_-)/\hbox{tr}\Omega(\sigma_+)
\!=\!\hbox{tr}\Omega(\sigma_-)/[\hbox{tr}\Omega(\sigma)\!+\!
\hbox{tr}\Omega(\sigma_-)]$, we have $0\!\le\!M_2^{sep}(\sigma)\!\le\!1$.
Likewise to the case of $M_1^{sep}(\sigma)$, the dual problem for
$M_2^{sep}(\sigma)$ is 
\begin{eqnarray}
&& M_2^{sep}(\sigma)=\min_{\sigma_\pm}\min x,\cr
&&\sigma-(1-x)\sigma_+ \hbox{~is 1-positive}\cr
&&\sigma=\sigma_+-\sigma_-,\,\,\,\,\,\,\sigma_\pm\ge0,\,\,\,\,\,\,
\hbox{$\sigma_+$ is searable.}
\label{eq: monotone 2 dual}
\end{eqnarray}
\par
%
For every separable $\sigma$,
$M_2^{sep}(\sigma)\!=\!0$ 
because the decomposition of $\sigma_+\!=\!\sigma$ and $\sigma_-\!=\!0$ exists.
On the other hand, it is somewhat surprising that
$M_2^{sep}(|\psi\rangle)\!=\!1$ for every entangled $|\psi\rangle$.
To see this, let $\sigma\!=\!P_2^+$ on ${\mathbb C}^2\!\otimes\!{\mathbb C}^2$,
and let $Z\!\equiv\!P_2^+\!-\!(1\!-\!x)\sigma_+\!=\!(F\!\otimes\!\Theta)P_2^+$
using a local filtering map $F$ and some map $\Theta$.
When $F$ corresponds to a rank-1 projection,
$Z$ is 1-positive only when $Z\!\ge\!0$.
However, $Z\!\ge\!0$ for $x\!<\!1$ only when $\sigma_+\!=\!P_2^+$,
which is not separable and not allowed as a decomposition of $\sigma$.
When $F$ corresponds to a projection of rank-2,
$Z$ is 1-positive if and only if $Z$ is decomposable, and hence
$Z\!=\!X+Y^\Gamma$ must hold with $X,Y\!\ge\!0$.
For $x\!<\!1$, $Z$ is not positive, and hence $Y^\Gamma$ must have a negative
eigenvalue, $Y^\Gamma$ must have three positive
(and one negative) eigenvalues \cite{Verstraete01d,Ishizaka04a},
and $Z\!=\!X\!+\!Y^\Gamma$ must have three
positive eigenvalues.
However, $Z$ cannot have three positive eigenvalues because of
$\sigma_+\!\ge\!0$.
After all, the constraints in Eq.\ (\ref{eq: monotone 2 dual}) cannot be
satisfied for $x\!<\!1$, and hence $M_2^{sep}(P_2^+)\!=\!1$.
Since all pure entangled $|\psi\rangle$ can be converted to $P_2^+$ by LOCC
and $M_2^{sep}$ never increases under LOCC, $M_2^{sep}(|\psi\rangle)\!=\!1$.
Moreover, as in the case of the example (iv) for $M_1(\sigma)$,
for the mixed state of 
\begin{equation}
\sigma=\sigma_0+\lambda\sigma_+^*,
\end{equation}
where $\sigma_+^*$ constitutes an optimal decomposition for $\sigma_0$,
it is found that 
\begin{equation}
M_2^{sep}(\sigma)\le\frac{M_2^{sep}(\sigma_0)}{1+\lambda},
\end{equation}
since $(1\!+\!\lambda)\sigma_+^*$ is not necessarily optimal for $\sigma$.
In this way, $M_2^{sep}(\sigma)\!\in\![0,1]$ certainly represents 
non-trivial strong monotonicity under LOCC, and we have
\par
{Theorem 2:}
{\it
The function $M_2^{sep}(\sigma)$, which is defined by
Eq.\ (\ref{eq: monotone 2 dual}), is a strong monotone under LOCC.
If $\sigma$ is single-copy distillable under LOCC, then
$M_2^{sep}(\sigma)\!=\!1$.
}
\par
%
Note that 
$M_2^{sep}(\sigma)$ and $M_2(\sigma)$ (defined below) are
lower bounded by the generalized robustness of entanglement $R_G$
\cite{Steiner03a} as
\begin{equation}
M_2(\sigma)\ge M_2^{sep}(\sigma)\ge\min_{\sigma_\pm}\frac{\hbox{tr}\sigma_-}{\hbox{tr}\sigma_+}
=\frac{R_G(\sigma)}{1+R_G(\sigma)}.
\end{equation}
Here, $M_2(\sigma)$ is a strong monotone for which ${\cal L}_\sigma$
in Eq.\ (\ref{eq: monotone 2}) is chosen as the set of PPT-preserving
operations, and hence $M_2(\sigma)$ is defined such that the constraint of
the 1-positivity in Eq.\ (\ref{eq: monotone 2 dual}) is replaced by the
constraint of the decomposability.
\par
%
\section{Summary}
\label{sec: Summary}
%
We proposed four strong entanglement monotones, $M_1$, $M_1^{sep}$,
$M_2$, and $M_2^{sep}$, and studied those properties.
All these strong monotones take 1 for every pure entangled states and
hence represent the strong monotonicity independent on the Schmidt number in
mixed-state entanglement manipulation.
Moreover, these strong monotones provide necessary conditions for
single-copy distillability.
These are lower bounded by the negativity or generalized
robustness of entanglement.
All these strong monotones are derived from the strong monotone function
$M(\sigma)$ given by Eq.\ (\ref{eq: function M}), whose optimization problem
of the supremum concerning LOCC or PPT-preserving operations can be reduced
to a simple convex optimization problem.
The constraint of the dual optimization problem is described by the
decomposability and 1-positivity of an operator constructed from a given
quantum state, and it clearly reveals the geometrical characteristics of
entangled state as shown in Fig.\ \ref{fig: Geometrical structure}.
In this paper, we concentrated our attention on the bipartite systems
but it is obvious that these strong monotones can be applicable to
every bipartite partitioning of multipartite systems. 
\par
%
\par
Finally, let us briefly discuss the relation between the strong monotones
studied in this paper and asymptotic distillability.
If $\sigma$ is asymptotically distillable, there must exist a trace-preserving
LOCC (ended by twirling) which produces an isotropic state
close to a maximally entangled state,
namely there must exist LOCC $\Lambda$ such that 
$\Lambda(\sigma^{\otimes n})\!=\!\sigma_I$ with 
$\eta\!\rightarrow\!1$ for $n\!\rightarrow\!\infty$.
Here $\sigma_I$ is an isotropic state given by Eq.\ (\ref{eq: Isotropic state})
and  the distillable entanglement $E_D$ is given by
$(\log_2 d)/n\!\rightarrow\!E_D(\sigma)$ with $d$ being the dimension of
$\sigma_I$ \cite{Horodecki00a}.
On the other hand, 
$M_1(\sigma_I)\!\rightarrow\!1$ for $\eta\!\rightarrow\!1$ as
explicitly shown in Eq.\ (\ref{eq: monotone for an isotropic state})
[$M_1(\sigma)$ is a discontinuous function but it is continuous
on an isotropic state].
This implies that
\begin{equation}
1\ge M_1(\sigma^{\otimes n})\ge M_1(\Lambda(\sigma^{\otimes n})) = M_1(\sigma_I) \rightarrow 1,
\end{equation}
and therefore $M_1(\sigma^{\otimes n})$ must satisfy
$M_1(\sigma^{\otimes n})\!\rightarrow\!1$ for $n\!\rightarrow\!\infty$
so that $\sigma$ is asymptotically distillable.
In this way, the asymptotic behavior of $M_1(\sigma^{\otimes n})$
provides a condition necessary to the asymptotic distillability.
However, since the negativity satisfies 
$N(\sigma^{\otimes n})\!\rightarrow\!\infty$ for every non-PPT (NPT) states,
it is found using Eq.\ (\ref{eq: Lower bound for M_1}) that
$M_1(\sigma^{\otimes n})\!\rightarrow\!1$ for every NPT states.
Therefore, $M_1$ does not provide any non-trivial result concerning
asymptotic distillability unfortunately.
This is the case for $M_1^{sep}(\sigma^{\otimes n})$ also.
On the other hand, it is open whether
$M_2^{sep}(\sigma^{\otimes n})\!\rightarrow\!1$ for every NPT states or not
[as far as we know the asymptotic behavior of $R_G(\sigma^{\otimes n})$
for every NPT states has not been shown yet].
Note that the continuity of $M_2^{sep}$ on an isotropic state is also open.
\par
%
The problem obtaining the tractable criterion for single-copy distillability
in higher dimensional systems is still open as well as the asymptotic
distillability.
We wish the results in this paper could shed some light on these open problems.
\par
%
\begin{acknowledgements}
The author would like to thank M. Owari for helpful discussions.
\end{acknowledgements}
%

\begin{thebibliography}{10}

\bibitem{Yang05a}
D. Yang, M. Horodecki, R. Horodecki, and B. Synak-Radtke, Phys. Rev. Lett. {\bf
  95},  190501  (2005).

\bibitem{Vidal00a}
G. Vidal, J. Mod. Opt. {\bf 47},  355  (2000).

\bibitem{Bennett96a}
C.~H. Bennett, D.~P. DiVincenzo, J.~A. Smolin, and W.~K. Wootters, Phys. Rev. A
  {\bf 54},  3824  (1996).

\bibitem{Vedral97a}
V. Vedral, M.~B. Plenio, M.~A. Rippin, and P.~L. Knight, Phys. Rev. Lett. {\bf
  78},  2275  (1997).

\bibitem{Vidal02a}
G. Vidal and R.~F. Werner, Phys. Rev. A {\bf 65},  032314  (2002).

\bibitem{Plenio05a}
M.~B. Plenio, Phys. Rev. Lett. {\bf 95},  090503  (2005).

\bibitem{Vidal99b}
G. Vidal and R. Tarrach, Phys. Rev. A {\bf 59},  141  (1999).

\bibitem{Steiner03a}
M. Steiner, Phys. Rev. A {\bf 67},  054305  (2003).

\bibitem{Karnas01a}
S. Karnas and M. Lewenstein, J. Phys. A {\bf 34},  6919  (2001).

\bibitem{Lewenstein98a}
M. Lewenstein and A. Sanpera, Phys. Rev. Lett. {\bf 80},  2261  (1998).

\bibitem{PlenioVirmani05a}
M.~B. Plenio and S. Virmani, quant-ph/0504163.

\bibitem{Terhal00b}
B.~M. Terhal and P. Horodecki, Phys. Rev. A {\bf 61},  040301(R)  (2000).

\bibitem{Vidal99a}
G. Vidal, Phys. Rev. Lett. {\bf 83},  1046  (1999).

\bibitem{Dur00b}
W. D{\" u}r, G. Vidal, and J.~I. Cirac, Phys. Rev. A {\bf 62},  062314  (2000).

\bibitem{Ishizaka04b}
S. Ishizaka, Phys. Rev. Lett. {\bf 93},  190501  (2004).

\bibitem{Ishizaka05a}
S. Ishizaka and M.~B. Plenio, Phys. Rev. A {\bf 71},  052303  (2005).

\bibitem{Rains99b}
E.~M. Rains, Phys. Rev. A {\bf 60},  173  (1999).

\bibitem{Peres96a}
A. Peres, Phys. Rev. Lett. {\bf 77},  1413  (1996).

\bibitem{Horodecki96a}
M. Horodecki, P. Horodecki, and R. Horodecki, Phys. Lett. A {\bf 223},  1
  (1996).

\bibitem{Lewenstein00a}
M. Lewenstein, B. Kraus, J.~I. Cirac, and P. Horodecki, Phys. Rev. A {\bf 62},
  52310  (2000).

\bibitem{Sanpera01a}
A. Sanpera, D. Bru{\ss}, and M. Lewenstein, Phys. Rev. A {\bf 63},  050301(R)
  (2001).

\bibitem{Kraus02a}
B. Kraus, M. Lewenstein, and J.~I. Cirac, Phys. Rev. A {\bf 65},  042327
  (2002).

\bibitem{Ha03a}
K.-C. Ha, S.-H. Kye, and Y.~S. Park, Phys. Lett. A {\bf 313},  163  (2003).

\bibitem{Clarisse05a}
L. Clarisse, Phys. Rev. A {\bf 71},  032332  (2005).

\bibitem{Masanes05a}
L. Masanes, quant-ph/0508071  .

\bibitem{Wootters98a}
W.~K. Wootters, Phys. Rev. Lett. {\bf 80},  2245  (1998).

\bibitem{Stormer63a}
E. St{\o}rmer, Acta. Math. {\bf 110},  233  (1963).

\bibitem{Woronowicz76a}
S.~L. Woronowicz, Rep. Math. Phys. {\bf 10},  165  (1976).

\bibitem{Verstraete03b}
F. Verstraete and H. Verschelde, Phys. Rev. Lett. {\bf 90},  097901  (2003).

\bibitem{Verstraete01a}
F. Verstraete, J. Dehaene, and B. DeMoor, Phys. Rev. A {\bf 64},  010101(R)
  (2001).

\bibitem{Kent98a}
A. Kent, Phys. Rev. Lett. {\bf 81},  2839  (1998).

\bibitem{Horodecki99a}
M. Horodecki, P. Horodecki, and R. Horodecki, Phys. Rev. A {\bf 60},  1888
  (1999).

\bibitem{Boyd04a}
S. Boyd and L. Vandenberghe, {\em Convex Optimization} (Cambridge University
  Press, Cambridge, UK, 2004).

\bibitem{Cirac01a}
J. I. Cirac, W. D{\" u}r, B. Kraus, and M. Lewenstein,
Phys. Rev. Lett. {\bf 86}, 544 (2001).

\bibitem{Brandao05a}
F.~G. S. L.~Brand{\~a}o, Phys. Rev. A {\bf 72},  022310  (2005).

\bibitem{Verstraete01d}
F. Verstraete, K. Audenaert, J. Dehaene, and B.~D. Moor, J. Phys. A {\bf 34},
  10327  (2001).

\bibitem{Ishizaka04a}
S. Ishizaka, Phys. Rev. A {\bf 69},  020301(R)  (2004).

\bibitem{Horodecki00a}
M. Horodecki, P. Horodecki, and R. Horodecki, Phys. Rev. Lett. {\bf 84},  2014  (2000).

\end{thebibliography}

%
\end{document}